\def\year{2025}\relax
\documentclass[letterpaper]{article} 
\usepackage{aaai25}  
\usepackage{times}  
\usepackage{helvet}  
\usepackage{courier}  
\usepackage[hyphens]{url}  
\usepackage{graphicx} 
\usepackage{natbib}  
\usepackage{caption} 
\DeclareCaptionStyle{ruled}{labelfont=normalfont,labelsep=colon,strut=off} 
\usepackage{algorithm}
\usepackage{algorithmic}

\usepackage{booktabs}
\usepackage{multirow}
\usepackage{float}
\usepackage{xcolor}
\newcommand{\answerYes}[1]{\textcolor{blue}{#1}} 
\newcommand{\answerNo}[1]{\textcolor{teal}{#1}} 
\newcommand{\answerNA}[1]{\textcolor{gray}{#1}} 
 
\usepackage{newfloat}
\usepackage{listings}
\floatstyle{ruled}
\newfloat{listing}{tb}{lst}{}
\floatname{listing}{Listing}
\title{Differentiating Emigration from Return Migration of Scholars Using Name-Based Nationality Detection Models}

\author {
    Faeze Ghorbanpour\textsuperscript{\rm 1,3,4}\thanks{This work was conducted while the author was visiting the Max Planck Institute for Demographic Research.},
    Thiago Zordan Malaguth\textsuperscript{\rm 2},
    Aliakbar Akbaritabar\textsuperscript{\rm 2}
}
\affiliations {
    \textsuperscript{\rm 1}School of Computation, Information and Technology, TU Munich\\
    \textsuperscript{\rm 2}Max Planck Institute for Demographic Research, Rostock, Germany\\
    \textsuperscript{\rm 3}Center for Information and Language Processing, LMU Munich\\
    \textsuperscript{\rm 4}Munich Center for Machine Learning (MCML)\\
    faeze.ghorbanpour@tum.de, zordanmalaguth@demogr.mpg.de, akbaritabar@demogr.mpg.de
}

\begin{document}

\maketitle

\begin{abstract}

Most web and digital trace data do not include information about an individual's nationality due to privacy concerns. The lack of data on nationality can create challenges for migration research. It can lead to a left-censoring issue since we are uncertain about the migrant's country of origin. Once we observe an emigration event, if we know the nationality, we can differentiate it from return migration. We propose methods to detect the nationality with the least available data, i.e., full names. We use the detected nationality in comparison with the country of academic origin, which is a common approach in studying the migration of researchers. We gathered 2.6 million unique name-nationality pairs from Wikipedia and categorized them into families of nationalities with three granularity levels to use as our training data. Using a character-based machine learning model, we achieved a weighted F1 score of 84\% for the broadest- and 67\%, for the most granular, country-level categorization. In our empirical study, we used the trained and tested model to assign nationality to 8+ million scholars' full names in Scopus data. Our results show that using the country of first publication as a proxy for nationality underestimates the size of return flows, especially for countries with a more diverse academic workforce, such as the USA, Australia, and Canada. We found that around 48\% of emigration from the USA was return migration once we used the country of name origin in contrast to 33\% based on academic origin. In the most recent period, 79\% of scholars whose affiliation has consistently changed from the USA to China, and are considered emigrants, have Chinese names in contrast to 41\% with a Chinese academic origin. Our proposed methods in addressing left-censoring issues are beneficial for other research that uses digital trace data to study migration.
\end{abstract}

\section{Introduction}
Sociological and demographic research has increasingly used web and digital trace data \cite{kashyap2023DigitalComputationalDemography,alburez-gutierrezDemographyDigitalEra2019}. 
The nationality of an individual is not included in most digital trace data to preserve privacy. If we only use observational data for migration research, we cannot be certain about the country of origin of a migrant.
Hence, the migration trajectories based on the digital traces could be prone to a left-censoring issue because we do not know if the first emigration event observed in digital traces is from the actual country of origin. This emigration event could have been recorded in one of the next locations along the migration trajectory. To address this, we need to identify an individual's nationality. Here, we proposed methods for this identification that use the least available information, i.e., full names. If we accurately identify nationalities using names, we can complement observational data by adding a baseline for comparison to resolve uncertainties in migration trajectories.

Bibliometric data, which is information extracted from scientific publications, is increasingly used for demographic research \cite{kashyap2023DigitalComputationalDemography}. The literature highlights its value as a resource for analyzing the migration patterns of actively publishing researchers by examining consistent changes in the countries of affiliation \cite{zhao2023GenderPerspectiveGlobal,akbaritabar2024BilateralFlowsRates,sanliturk2023global}. Bibliometric databases do not provide information about a scholar's country of origin or nationality. As a result, researchers use various methods and data to determine the country of origin for these individuals, such as census data \cite{lewison2016lung, grilli2017last}. Using the country of affiliation where a scientist published their first paper as their country of ``academic birth" is a common practice \cite{kashyap2023DigitalComputationalDemography,zhao2023GenderPerspectiveGlobal,thelwall2023can}. Here, we aimed to provide an alternative method for identifying scholars' nationalities through bibliometric data by using their names. By recognizing the nationalities of scholars, we can re-identify the composition of the population of scholars per country. We can recalculate migration rates with this alternative measurement of the country of origin. Furthermore, the detected nationality helps differentiate emigration from \textit{return migration} of the graduate students to their home countries who were identified as emigrants based on their academic origin  \cite{sanliturk2023global}.

Names are often culturally specific and passed down through generations. They may reflect the nationality, race, or ethnicity of one's parents. Consequently, full names can sometimes serve as indicators to infer these attributes \cite{newman2018name, schaefer2013relationship}. 
However, we acknowledge that identifying someone's nationality, race, or ethnicity, based solely on their name is not always reliable, particularly among second-generation migrants, children of mixed-ethnic families, and those who experience changes in names. Additionally, the nationality identified using names could not be considered an equivalent for citizenship \citep{amit2023you, vathi2015identities}. Nevertheless, in the context of migration research, when we have already observed a consistent movement event, if we identify the nationality using names, we can differentiate whether this is more likely to be an emigration or a return migration. 
Language models have shown remarkable proficiency in text-based tasks, including identifying patterns in names to predict nationality. ByT5 models are considered one of the most advanced language models for character-based text processing \cite{xue-etal-2022-byt5}.

This paper aims to differentiate between the emigration and return migration of actively publishing researchers using Scopus bibliometric data, comprising 28+ million publications from 8.2 million researchers between 1996 and 2020. The dataset includes the names and affiliation addresses of scholars but lacks information on the countries of birth or nationalities.
We use character-based language models to predict the likely origin of scholars' names. We train these models using 2.6 million unique name-nationality pairs from individuals with Wikipedia pages, which is independent of Scopus data. Our training data is more comprehensive than prior studies that relied on publicly available data. Following the methodology of \citet{ye2017nationality} and \citet{ye2019secret}, we categorized nationalities into three hierarchical levels and customized them based on our training data, ranging from broad nationality or ethnic groups (12 classes) to more specific countries (175 classes).
After analyzing the composition of selected countries based on the country of name origin, we will explore the impact of using the origin of scholars' names as a proxy for their nationality. Specifically, we aim to assess whether this approach would alter the size and direction of return migration flows based on academic origin.
To support replication and further research, we publicly released the pre-trained models, country categorizations, and training data developed in this study.
The next sections are structured as follows: after reviewing the related works, we describe nationality detection, which covers data processing, categorization, model training, and evaluation. Following this, there is a section on our empirical study of Scopus bibliometric data that provides insights into the emigration and return migration of researchers.

\section{Related Work}

Bibliometric data is increasingly used to analyze scholars' migration. These studies consider consistent changes in affiliation addresses spanning more than one year as an indicator of a migration event. They used the country where a scientist published their first paper as the country of academic birth \cite{kashyap2023DigitalComputationalDemography,zhao2023GenderPerspectiveGlobal,sanliturk2023global,akbaritabar2024BilateralFlowsRates}. \citet{thelwall2023can} used academic origins and studied the uniqueness and distribution of first and last names. \citet{sanliturk2023global}, and \citet{akbaritabar2024BilateralFlowsRates} used academic origin to study the relationship between scholars' migration and economic development, and to investigate the global bilateral flows and rates of international migration, respectively. It is shown that a significant proportion of researchers move from developing to developed countries, a phenomenon often described as ``brain drain" \cite{dodani2005brain}. However, evidence of return migration suggests a ``brain circulation," where scholars contribute to their home countries after gaining international experience \cite{crescenzi2017they}. The literature also highlights the role of collaborative networks and institutional policies in facilitating or hindering scholars' mobility \cite{moed2013studying,walk2024ResolvingParadoxMigration}.

Recent studies have explored the return migration of scholars and the factors influencing it. \citet{zhao2022return} used Scopus data from 1996 to 2020 to analyze the departure and return of German-affiliated researchers ---based on country of academic origin--- and found that gender, academic cohort, and discipline significantly impact return migration. Male researchers were more likely to return to Germany compared to their female counterparts. Similar patterns were observed in a study on the return migration of academics by \citet{gaule2014comes}, which identified career opportunities, personal reasons, and institutional support as key drivers for scientists deciding when and if to return to their home country.

The literature also highlights the complexity of migration decisions beyond simple return scenarios. For example, scholars may engage in circular or onward migration (similar to migration pathways proposed by \citet{king2010jems}), where they continue to move between different countries rather than permanently returning to their home country, which could be driven by the global dynamics of academic opportunities and knowledge exchange \cite{constant2021return, constant2020time}. Hence, it is important to consider individual and contextual factors in the study of global academic mobility \citep{cattaneo2020multiple}. Furthermore, \citet{zhao2023GenderPerspectiveGlobal} have reported gender disparities in scholarly migration and showed that female researchers are less likely to migrate internationally, which could often be due to institutional, social, or familial constraints. It is important to consider demographic factors' effect on academic mobility. 

Researchers studying the distribution of scholars' nationalities and migration often use census data to determine the nationality. Census data provides the most common or popular names for each country. For example, this approach is used by \citet{lewison2016lung} to determine the nationality of lung cancer researchers based on the countries where their last names are most prevalent. A similar approach is taken by \citet{grilli2017last} for analyzing mobility, gender imbalance, and nepotism within academic systems. \citet{ioannidis2021comprehensive} examine scientists with common Greek surnames to identify Greek scientists in Greece and in the diaspora. 

In recent years, several methods have been developed to analyze the text of an author's name and estimate their nationality. For instance, \citet{le2021analysis} conducted an analysis of disparities in scientific society honors using machine learning to determine the origins of authors' names, and \citet{Jain2022Importance} used machine learning to predict ethnicity to analyze political donations from various ethnic groups. However, they applied a machine learning method to a smaller amount of training data.

Machine learning (ML) models have demonstrated exceptional performance in text-based tasks, such as identifying patterns in names to predict nationality, ethnicity, or race. Numerous studies have used ML for name-based classification. For instance, \citet{treeratpituk2012name} developed a name-ethnicity classifier using multinomial logistic regression, while \citet{lee2017name},  \citet{le2021analysis}, and \citet{Jain2022Importance} used recurrent neural networks, and \citet{kang2020name, chaturvedi2024s} used convolutional neural networks for name-nationality classification. These approaches, however, are often constrained by limited training datasets, leading to potential biases in the results. To mitigate such biases, it is crucial to train models on large datasets and evaluate them on distinct data from the training set. One example of a system addressing this is NamePrism \citep{ye2017nationality, ye2019secret}, which uses Naive Bayes to create name embeddings for classifying nationalities and ethnicities, revealing relationships between names across different groups. Despite its strength in leveraging extensive data, the proprietary nature of its dataset and lack of publicly accessible code restricts further exploration of its method by other researchers.

Following the presented literature, we aimed to provide an alternative method for identifying scholars' country of origin through bibliometric data by assigning a nationality using ML models. Our analysis uses authors' first and last names and a substantially larger amount of training data than previous research. We ask whether part of the high emigration rates reported in the literature \citep{akbaritabar2024BilateralFlowsRates,sanliturk2023global,zhao2023GenderPerspectiveGlobal}, from countries such as the USA to destinations such as China, could be the return migration of graduate students and scholars to their country of origin.

\section{Data and Methods}

To develop an ML model to assign a country of origin to names, we need a training dataset containing pairs of names and nationalities. We used this to train our classification models. The full names served as our textual data, while the nationalities served as the classes.
Obtaining such data proved challenging due to the private nature of personally identifiable information. 

Public Wikipedia introduction pages for individuals are a valuable resource as these pages often include a biography outlining the nationality, birthplace, or citizenship details. We gathered publicly available datasets based on Wikipedia, typically created for other purposes like text generation for ML. We combined seven different datasets, resulting in 2.6 million unique name and nationality pairs for training our classification models (see Table \ref{tab:datasets}). Additionally, we used DBpedia's Application Programming Interface (API)\footnote{\url{https://dbpedia.org/page/Web_API}} to fill in missing information from Wikipedia. We focused on pages that provided at least one of the following information: nationality, birthplace, or citizenship. 

We evaluated our model's performance on two test datasets from different sources than the Wikipedia dataset used for training (see the bottom rows of Table \ref{tab:datasets}, also see more information about data collection and pre-processing in Appendix). 

\begin{table*}[!htp]\centering
\scriptsize
\begin{tabular}{lp{5.5cm}p{8cm}l}\toprule
\textbf{} &\textbf{Name} &\textbf{Description} &\textbf{Volume} \\\midrule
\multirow{7}{*}{\textbf{Training data}} &Wikipedia notable people's mobility \cite{wiki-notable} &This dataset contains mobility data of notable people whose biography is registered in Wikipedia. &2,190,457 \\\cmidrule(lr){2-4}
&Nationality Estimate \cite{le2021analysis}  &This dataset is constructed with name-nationality pairs by parsing the English-language Wikipedia. &696,136 \\\cmidrule(lr){2-4}
&WikiBio (Wikipedia Biography Dataset) \cite{wiki-bio} &This dataset gathers biographies from English Wikipedia. &418,978 \\\cmidrule(lr){2-4}
&Wikipedia Person Dataset \cite{wiki-person}   &This dataset gathers the infobox with the description of the person's page. &355,468 \\\cmidrule(lr){2-4}
&Politicians on Wikipedia \cite{politicians} &This dataset contains information about politicians from Wikipedia. &54,877 \\\cmidrule(lr){2-4}
&Pantheon 1.0 dataset \cite{wiki-people} &This dataset includes the biographies present in more than 25 languages in Wikipedia. &16,071 \\\cmidrule(lr){2-4}
&Wikidata &Processed by our colleague. &131,731 \\\midrule
& &SUM &3,874,776 \\
& &\textbf{SUM after removing duplicates} &2,634,445 \\\midrule
& &\textbf{SUM after removing multi-cultural countries and small classes} &2,245, 768 \\\midrule 
\multirow{2}{*}{\textbf{Test data}} &Athletes \cite{kang2020name} &Data collected from the sports sites. &42,413 \\\cmidrule(lr){2-4}
&International Union for the Scientific Study of Population (IUSSP) public members' directory\footnote{\url{https://iussp.org/en/directorysearch}} &Information on IUSSP member's self-reported nationality, field of specialty, spoken languages, affiliation, etc. &1,547 \\
\bottomrule
\end{tabular}
\caption{Datasets used for training (top) and testing (bottom) the ML models.}\label{tab:datasets}
\end{table*}

We developed a hierarchical nationality categorization framework with three levels of granularity. Each level builds on the lower level, allowing for a clearer understanding of name similarities among individuals. The first level has 12 classes, the second has 30, and the third, country level, has 175. To do this, we used the frameworks developed by \citet{ye2017nationality} and \citet{le2021analysis} and adjusted them to fit our dataset. For instance, we added countries from Oceania and the Caribbean. This ensures a more comprehensive representation of naming patterns across diverse regions. We eliminated countries where we had fewer than 100 instances, as they were considered small classes. We also excluded countries with multicultural populations, such as the US, Canada, Australia, New Zealand, and South Africa, from the training and testing. However, consistent with prior frameworks, we included them in the empirical analysis, i.e., to assign nationality to names from Scopus data. In addition, we present comparative results on the composition of scholars in these countries of immigration to show their diverse scientific workforce. The UK, while a multicultural country, was chosen to represent English-origin names because it strikes a balance between diversity and classification performance, as reflected in its relatively lower entropy and higher F1 score compared to other predominantly English-speaking nations. For the rationale regarding the inclusion and exclusion of countries, please refer to Appendix.

\begin{figure}[!ht]
    \centering
    \includegraphics[width=\columnwidth]{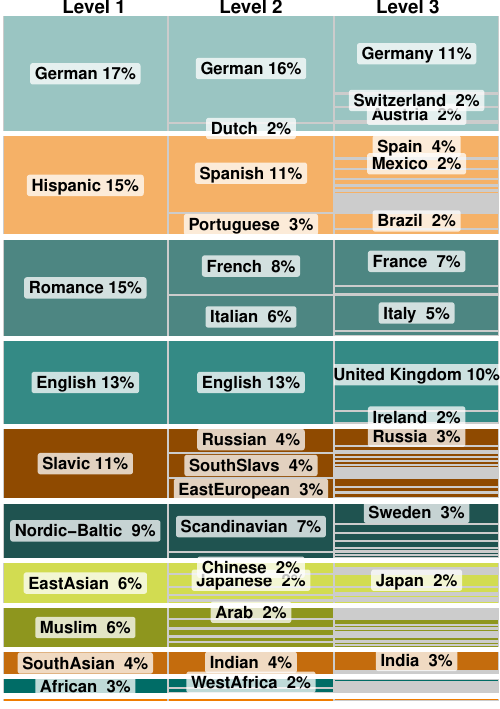}
    \caption{Distribution of nationality classes in the training dataset. Each column sums up to 100\% and represents one of the three classification levels, from the left (broader) to the right (more detailed). All classes are shown, but only those bigger than 1.5\% of the total are labeled.}
    \label{fig:treemap}
\end{figure}

Figure \ref{fig:treemap} presents our classification as a treemap, where the first column shows broader first-level groupings (e.g., ``German" and ``Dutch" as ``German"), the middle column groups the countries by second-level classes (e.g., Germany, Austria, and Switzerland as ``German"), and the third column corresponds to the main classification at the country-, i.e., most granular, level. The classes are ordered by size within each category. This figure also shows the distribution of classes in the training dataset at each of the three levels. At the country level, Germany is the largest class, followed by the UK, France, and Italy. At the second level, the German class is predominant, followed by English, Spanish, and French. At the broadest level, Germanic leads, followed by Hispanic, Romance, English, and Slavic. Although class sizes are influenced by the number of Wikipedia pages per country, we mitigated this by collecting more data and adjusting our methods to balance small and large classes. 

We also used two separate test sets to test the robustness of our trained models (see the bottom rows of Table \ref{tab:datasets}). It is important to note that these test sets have different class distributions; for example, in the IUSSP data, the major classes are India, Brazil, France, and Nigeria, and in the athletes' data, the largest classes are Indonesia, South Korea, and China  (See figures with the composition of classes in the Appendix). 

Letter patterns are important in full-name processing. We used character-based ML methods to detect these patterns. We used three neural network approaches. First, we used FastText, a skip-gram model that represents each word as a collection of character n-grams \cite{fasttext}. It is well-suited for name processing due to its ability to capture subword information, making it robust to rare or unseen names. Second, we used a character-based convolutional neural network (char-CNN), which includes an embedding layer, k CNN layers (with k as a hyperparameter), and layer normalization \cite{character-cnn, wu2017introduction}. CNNs are effective for processing names because they excel at capturing local patterns within character sequences, which is essential for distinguishing variations in names. Similarly, a character-based long short-term memory (char-LSTM), consisting of an embedding layer, k BiLSTM layers, and layer normalization, is particularly suited for handling sequential data like names \cite{verwimp2017character, graves2012long}. It captures long-range dependencies between characters, enabling better handling of name structure and variations. The best practices that are publicly available, and used in previous studies, include char-LSTM by \citet{lee2017name}, \citet{le2021analysis}, and \citet{Jain2022Importance}, and char-CNN by \citet{kang2020name} and \citet{chaturvedi2024s}.

Third, we used a byte-level transformer model based on the T5 architecture (ByT5) in three versions: small, base, and large \cite{xue-etal-2022-byt5}. ByT5 processes text at the byte level, making it inherently character-based, which is crucial for handling names with diverse structures and spellings. 
ByT5's character-level processing enables it to learn subtle orthographic and phonetic patterns that are often specific to different nationalities. This capability is particularly advantageous in name classification tasks \citep{lee2017name}. Furthermore, ByT5's token-free design enables it to handle noisy inputs and to generalize to rare or previously unseen names, which are common challenges in name-based classification tasks. \citet{huang-etal-2023-inducing} used ByT5 to study character representations within subword-based language models, and \citet{stankevivcius2022correcting} have used ByT5 for tasks such as diacritic restoration and typo correction.
We trained all three versions from scratch on our Wikipedia dataset without altering the model architecture.

To test the performance of our models, we divided our dataset into training, validation, and test subsets, comprising 65\%, 15\%, and 20\% of the data, respectively (see Table \ref{tab:datasets}). Models were trained on the training set, fine-tuned using the validation set with Optuna \citep{akiba2019optuna}, and then retrained on the combined training and validation sets before being evaluated on the test set using 5 random seeds. Due to computational constraints, this process was limited for larger models like ByT5. We also assessed performance on two additional test datasets (athletes and IUSSP data). Given the data imbalance, we used weighted F1 score, macro F1 score, and accuracy as evaluation metrics. The weighted and macro F1 score balances precision and recall, emphasizing minority class performance while checking accuracy to provide a fair measure of overall performance. Please refer to Appendix for more information about the hyperparameters and training details.


Table \ref{tab:comparison_1} provides a comparison of the performance of the models on the test set derived from the Wikipedia dataset as our first model assessment. The results demonstrate that ByT5 models, on average, slightly outperform other character-based methods in terms of accuracy and F1 scores across all three levels of classification. Based on the evaluations presented in this table, we used the ByT5 model, which performed better. At the first level of classification, the model achieves an accuracy of approximately 84\%. At the second level, the accuracy remains high at around 81\%. However, as the number of classes increases, the model's performance decreases to about 68\% at the third level. Nevertheless, considering the complexity arising from the 175 classes, this performance can still be considered acceptable, although it could be improved in future research. 
Accuracy at the third level varies considerably between classes. Despite an average F1-weighted score of about 68\%, some countries, such as Japan, have F1 scores above 90\%, while others have F1 scores below 10. The four countries for which the model is most accurate belong to the ``East Asian" at the first level, while 6 of the 10 countries with the lowest F1 score belong to the Hispanic class (see Figure \ref{fig:evaluation_3} in Appendix). 

\begin{table*}[!htp]\centering
\scriptsize
\begin{tabular}{lcccccccccc}\toprule
&\multicolumn{3}{c}{Level 1} &\multicolumn{3}{c}{Level 2} &\multicolumn{3}{c}{Level 3} \\\cmidrule(lr){2-4}\cmidrule(lr){5-7}\cmidrule(lr){8-10}
&F1-macro &F1-weighted &Accuracy &F1-macro &F1-weighted &Accuracy &F1-macro &F1-weighted &Accuracy \\\midrule
Char-LSTM & 24.191 & 37.892 & 27.712 & 05.493 & 15.103 & 30.124 &  03.118 & 05.980& 16.536\\
Char-CNN &36.432 &53.929 &53.929 &53.929 &53.929 &53.929 &18.906 &06.785 &18.906 \\
Fasttext &\textbf{80.887} &83.598 &83.619 &69.132 &78.487 &78.956 &33.205 &63.445 &65.139 \\
ByT5 small &74.566 &\textbf{83.854} &\textbf{83.759} &\textbf{72.737} &80.971 &\textbf{81.121} &41.666 &\textbf{66.789} &\textbf{68.356} \\
ByT5 base &74.247 &83.557 &83.453 &72.491 & \textbf{80.673} &80.828 &\textbf{42.044} &66.453 &67.672 \\
ByT5 large &79.765 &82.987 &82.944 &72.051 &80.209 &80.349 &41.723 &65.799 &66.822 \\
\bottomrule
\end{tabular}
\caption{Evaluation results based on the Wikipedia test set.}\label{tab:comparison_1}
\end{table*}

These findings highlight the effectiveness of the ByT5 models in handling the classification task in comparison with other character-based neural networks. FastText outperforms both char-CNN and char-LSTM models. Additionally, since FastText's core code is written in C++ rather than Python, it is the fastest model among those we used (Refer to Appendix for more information).


\begin{table*}[!htp]\centering
\scriptsize
\begin{tabular}{lcccccccccc}\toprule
&\multicolumn{3}{c}{Level 1} &\multicolumn{3}{c}{Level 2} &\multicolumn{3}{c}{Level 3} \\\cmidrule(lr){2-4}\cmidrule(lr){5-7}\cmidrule(lr){8-10}
&F1-macro &F1-weighted &Accuracy &F1-macro &F1-weighted &Accuracy &F1-macro &F1-weighted &Accuracy \\\midrule
Athletes &79.276 &83.423 &83.222 &75.739 &79.889 &79.889 &40.916 &66.433 &66.003 \\
IUSSP &71.360 &84.669 &84.223 &74.597 &82.243 &82.119 &41.729 &66.523 &66.262 \\
\bottomrule
\end{tabular}
\caption{Evaluation results based on the two out-of-domain test datasets using the best-performing ByT5 models.}
\label{tab:comparison_2}
\end{table*}

As a second model assessment, we used two out-of-domain and separate test sets from different sources. While our model has not seen these datasets, and these data have different origins and distributions (see Figure \ref{fig:treemap_test} in Appendix), the metrics are almost the same as the performance on the Wikipedia test set (see Table \ref{tab:comparison_2}), demonstrating the robustness of the performance of the trained ByT5 models. The reasons for this are twofold: First, we used a sufficiently large training dataset to ensure that we have a reasonable number of samples, even for minor classes. Second, we used a character-based large language model (ByT5) that includes several layers of transformer and attention modules, which allows it to extract subtle and hidden patterns in names.

\section{Results}
We aimed to compare the emigration flows of researchers based on the country of academic origin ---country of affiliation in the first publication--- with nationality predicted using machine learning. We used a dataset from Scopus containing 8.2 million full author names and countries of affiliation throughout an author's career. 

\begin{figure*}[!ht]
    \centering
    \includegraphics{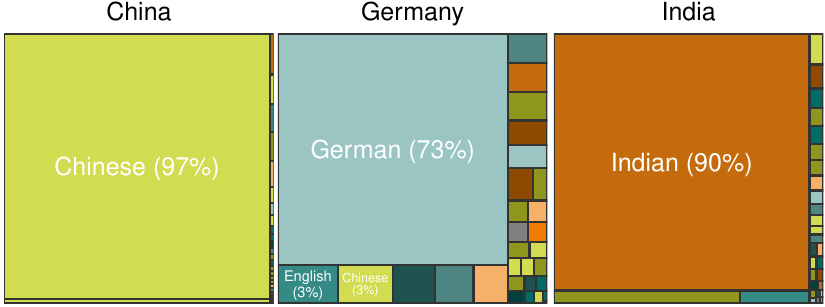}   
    \includegraphics{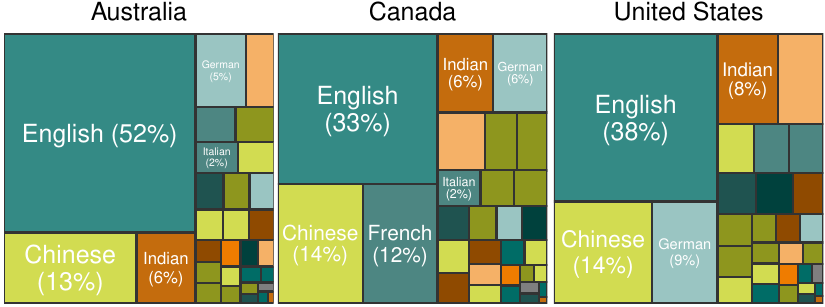}
    \caption{Composition of scholars affiliated with countries (names printed above panels) included in (top row) and excluded from the training data (bottom row) based on the country assigned using ML model (Level 2 regions, printed inside tree map).}
    \label{fig:composition_excluded}
\end{figure*}

First, we investigated the composition of a country's population of scholars based on ML-assigned nationalities (level 2 regions). The top row in Figure \ref{fig:composition_excluded} shows the three countries that had the largest populations in Scopus among those countries included in the training dataset. It shows that 97\% of scholars affiliated with institutions in China have Chinese names. 73\% of scholars in Germany have German names, and 90\% of scholars in India have Indian names. These percentages indicate that China has a smaller share of scholars with non-Chinese names, and Germany, in comparison to China and India, has a higher share of scholars with non-German names. Despite the difference between these three countries, they have a more homogeneous population compared to the countries excluded from the training, which are shown in the bottom row in Figure \ref{fig:composition_excluded}. These countries were excluded from training data due to their multicultural population and the figure shows our model's capability to identify the diversity of origin of names in these countries. Most authors in Australia, Canada, and the USA have English names with about 52\%, 33\%, and 38\%, respectively. In all these countries, the second most frequent origin of names is Chinese, with about 13\%, 14\%, and 14\%, respectively. The third most common origin differs between these countries, with Indian names being the third most common in Australia (6\%), and French (12\%) in Canada, probably due to the predominance of French-speaking individuals in Canada. In the USA, German names are the third most common origin. So far, our results do not differentiate between second-generation migrant scholars in a country who might have names that are different from the popular names in the host country, which will be discussed next.

\begin{figure*}[!ht]
    \centering
    \includegraphics{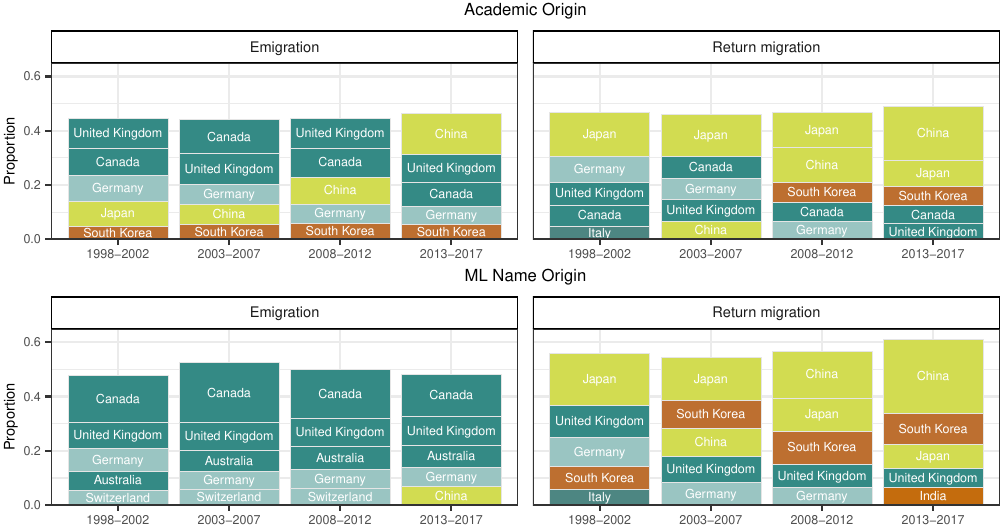}    
    \caption{Top 5 countries with the highest proportion of scholars among those affiliated to institutions in the United States divided into emigration (left bars) and return migration (right bars) based on academic origin (top bars) versus the country of ML-name origin (bottom bars).
    }
    \label{fig:bar_top5_destination_v3}
\end{figure*}

In the context of migration research, and once we have already observed a movement based on a consistent change in the affiliation addresses of scholars, we can use the ML nationality to further probe into aggregated emigration flows at the country level and ask if a subset of these emigrations could be considered returns to the country of origin. Here we wanted to identify the top destinations for scholars who emigrate versus those who return, based on their country of 1) academic origin and 2) ML-assigned name origin. 
We chose the USA for this analysis because it has the largest population of researchers in the Scopus database and is one of the countries with the most diverse population of researchers in terms of nationality (see Figure \ref{fig:composition_excluded}). 
From all scholars affiliated with the USA, Figure \ref{fig:bar_top5_destination_v3} represents the proportion of movements to the five most common destinations for emigration (left) and the five most common destinations for return movements (right) classified based on academic origin (top) and ML name origin (bottom). 
Based on academic origin, the most common emigration destinations from the USA were the UK, Canada, Germany, China, Japan, and South Korea. The most common destinations for return migration were similar but with differences in the order and size of the flows. 

ML name origin shows a different pattern and enables us to differentiate emigration from return migration. Using the ML name origin, Japan and South Korea are no longer among the top five most common destinations for emigration, and China appears only in the last period as the fifth most common destination; while it was the first destination based on the academic origin. This shift in the list of most common destinations for emigration and return migration suggests that using the academic origin might underestimate the size of return flows. When considering academic origin, the top 5 destinations concentrated less than 50\% of emigrations. However, when considering the ML-name origin, this tendency was more than 50\% of return movements for all considered time periods, with a higher share for East Asian and South Asian countries. These findings align with previous research indicating that Chinese and West Asian scholars have a higher tendency to return to their home countries after an international migration experience \citep{scab021, hu2021understanding}.

\begin{figure*}[!ht]
    \centering
  \includegraphics{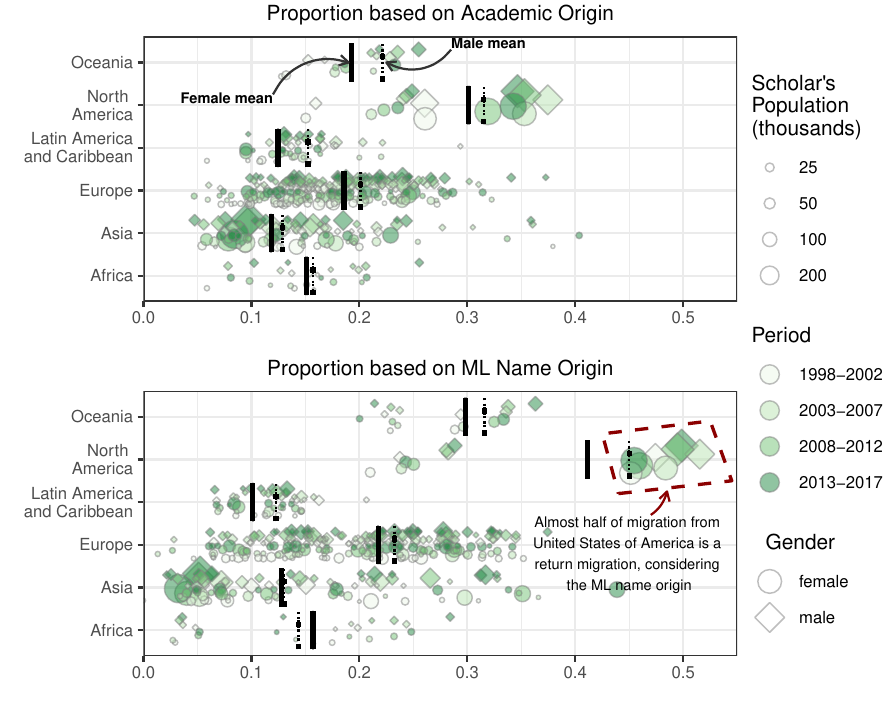}
    \caption{
    The proportion of return migration based on academic origin (top panel) versus the country of ML-name origin (bottom panel) divided by female scholars (diamonds) and male scholars (circles), relative to the size of the population of scholars (point size).}
    \label{fig:scatter-proportion}
\end{figure*}

Figure \ref{fig:scatter-proportion} shows the proportion of scholars who return to their home countries after migrating, compared to the total number of scholars who migrate across six continents and genders. To identify the gender of scholars, we used the method introduced by \citet{zhao2023GenderPerspectiveGlobal}. The mean proportion of return migration based on ML name origin was higher than that found based on academic origin (see the shift in points toward the right in the bottom panel compared to the top panel.) While similar patterns were observed across most continents, return migration proportions were much higher based on name origin for Oceania, North America, and Europe. Nearly half of the scholars leaving the USA return to their country of ML name origin. This indicates that these countries have a much higher share of foreign graduate students and scholars who start their publishing careers in these countries and later return to their countries of origin.

Based on the averages for men and women (vertical dashed and solid bars), male scholars tend to return to their home countries more often than female scholars, except in Africa, where the trend is reversed when considering ML name origin. This tendency is particularly strong for male scholars who were located in North America compared to scholars from other continents. This could signal differences in the rate of success in being hired permanently in the host country among genders and nationalities that require further research.

\section{Discussion and Conclusions}
Scholars contribute to knowledge production and innovation. Research has shown that migrant scholars are among those with the highest productivity and impact \cite{sugimoto2017ScientistsHaveMost}. Studying the migration of scholars, which can be traced using a consistent change in affiliation addresses in publication records \cite{kashyap2023DigitalComputationalDemography}, enables identifying  the countries that send scholars from those that receive them. However, without information about scholars' nationality or country of origin, it is difficult to interpret the migration patterns \cite{sanliturk2023global} due to left-censoring issues in digital trace data \cite{akbaritabar2024BilateralFlowsRates}. Particularly when assessing whether scholars are emigrating or returning to their country of origin.

This left-censoring problem could affect all migration research that relies solely on observational digital trace data. To address this problem, we used the most recent Machine Learning (ML) methods and language models to detect the nationality of scholars based on their full names. We used substantially larger data than previous studies to train models. We tested these models with various datasets and selected the best-performing ones. In our empirical study, and after testing our model's performance using different sets of test data, we used the model to assign names to 8.2 million scholars from Scopus bibliometric data. This enabled us to compare migration trends based on the country of academic origin and the country assigned using the ML model. It further allowed us to differentiate between emigration and return migration in the case of graduate students or more junior scholars who start their publication career in the host country and later return to their country of origin.

We observed substantial changes in emigration and return migration flows. For instance, the literature reports high rates of emigration from the USA to China (using the country of academic origin) \cite{zhao2023GenderPerspectiveGlobal,sanliturk2023global,akbaritabar2024BilateralFlowsRates}. Here, we asked whether part of this emigration could be the return migration of Chinese scholars who graduated in the USA and returned to their country of origin. It is evident that some of the scholars with Chinese names in the USA could be second generation migrants. Nevertheless, in the context of migration research, and when we have already observed a migration event, we can investigate with higher certainty if the movement is a return migration rather than emigration.

As we observed in the most recent period, 79\% of those emigrating from the USA to China have Chinese names and are categorized as returnees using the nationality assigned by machine learning in contrast to 41\% based on academic origin. 
Our observation aligns with China's government-initiated ``Thousand Talents Plan"\footnote{\url{https://en.wikipedia.org/wiki/Thousand_Talents_Plan}}, launched in 2008. This program aimed to attract top international talent to work in mainland China \citep{jia2018china} and is shown to have succeeded in motivating the return of Chinese scholars with international migration experience \citep{shi2023has}.
This could also be associated with some recent policies, such as the US-China Initiative. Launched in 2018 to counter economic espionage, this initiative has been criticized for disproportionately targeting Chinese-origin academics, creating a chilling effect on collaboration and increasing return migration to China \citep{jia2024impact}. Studies report a significant rise in Chinese-born scientists leaving the USA since the initiative's inception, with many returning to China \citep{xie2023CaughtCrossfireFears}. While the temporal scope of our data from 1998 to 2017 (we exclude two years at the start and end of the data from 1996 to 2020 to minimize left- and right censoring and truncation issues) is not exactly overlapping with this recent policy change, however, the literature emphasizes that scholars are among the subsets of the population that have credentials and networks to react faster to upcoming changes. 

We publicly share our model's code and training data to enable future research. Follow-up studies using digital trace data could use our model to extend the application of language models to analyze the migration patterns beyond the case of actively publishing researchers, for instance, to other data from social media, LinkedIn, etc. Furthermore, surveys of scholars' actual behavior could enable further validation of the model's ability to reliably predict scholars' migration trajectories, including their return or continued residence in host countries. Additionally, our models could be further refined to account for potential cultural, regional, gender, and disciplinary differences in migration behavior. Future research could investigate the possibility of long-term stays and integration in host countries that might be different between subsets of migrants depending on their nationalities and countries of origin.

To clarify the ethical terms and best practices in machine learning research, we emphasize that our models were trained using publicly available datasets that had been collected from Wikipedia. Our use of these datasets was solely for research purposes and completely complies with the licensing and use terms of this platform. Since the datasets only include names and nationalities of individuals from public Wikipedia pages, they do not contain private information.
We gain access to Scopus data through a license\footnote{Please see the acknowledgment section for details.} to do our empirical study on migration trends. This data is accumulated using publicly accessible information from scientific publications and does not constitute private information.

Our research has certain limitations that we would like to acknowledge. 
In preprocessing the data, we removed non-ASCII and redundant characters. However, this process could also eliminate information, such as language-specific umlauts and characters, that might be useful for classification. While we evaluated the effect of this preprocessing step, future work could conduct more experiments beyond the ones presented here to assess the impact of the preprocessing.
Additionally, ML and language models are valuable tools for detecting nationality, but they have uncertainty. While our model has a relatively high accuracy rate, this performance primarily applies to nationality groups with the highest representation in the training data. Predictions for nationality groups with a small number of observations have lower F1 scores (please see confusion matrices in the Appendix). While we use weighted measures to circumvent such problems, and these issues are less prevalent in migration corridors with the highest count of scholars such as those presented here, some errors can still occur and be explained by factors such as second-generation migrants, name or citizenship changes, or the diverse ways people choose names (please refer to the in the Appendix). To address this, we collected a large dataset and used one of the most advanced language models available. However, there are still errors, particularly for nationalities and groups with lower representation in training data, that remain unexplained. These inaccuracies can impact analyses of emigration and return migration among scholars and could be improved in future research.

Despite these limitations, our methods and publicly shared model could help reduce left-censoring issues in migration research using digital trace data. Our results highlight that a substantial part of the emigration, as reported in the literature using observational data, could be considered return migration, which has implications for migration research that uses digital trace data.

\section{Availability of Model Code and Training Data}
The training dataset used in this study is publicly available on Zenodo~\cite{ghorbanpour2025training}. 
The scripts to replicate our analysis and figures are accessible via the following GitHub repository:~\url{https://github.com/FaezeGhorbanpour/NameBasedNationalityDetection}.


\section{Acknowledgements}
We thank Tom Theile for lending us one of the datasets presented in the text based on Wikidata. This study has received access to the bibliometric data through the project ``Kompetenznetzwerk Bibliometrie'' \cite{schmidt2024DataInfrastructureGerman}, and we acknowledge their funder Bundesministerium für Bildung und Forschung (grant number 16WIK2101A).

\bibliography{aaai25}


\section{Ethics Checklist}

\begin{enumerate}

\item For most authors...
\begin{enumerate}
    \item  Would answering this research question advance science without violating social contracts, such as violating privacy norms, perpetuating unfair profiling, exacerbating the socio-economic divide, or implying disrespect to societies or cultures?
    \answerYes{Yes, see the Discussion.}
  \item Do your main claims in the abstract and introduction accurately reflect the paper's contributions and scope?
    \answerYes{Yes.}
   \item Do you clarify how the proposed methodological approach is appropriate for the claims made? 
    \answerYes{Yes, see the empirical study of migration patterns in Scopus Data.}
   \item Do you clarify what are possible artifacts in the data used, given population-specific distributions?
    \answerYes{Yes, see Appendix}
  \item Did you describe the limitations of your work?
    \answerYes{Yes, see the Discussion section.}
  \item Did you discuss any potential negative societal impacts of your work?
    \answerYes{We believe this work does not have negative societal impacts.}
      \item Did you discuss any potential misuse of your work?
    \answerYes{We believe this work can not be misused.}
    \item Did you describe steps taken to prevent or mitigate potential negative outcomes of the research, such as data and model documentation, data anonymization, responsible release, access control, and the reproducibility of findings?
    \answerYes{We will release the model documentation, data, and code for nationality detection and will explain how to reproduce the findings. }
  \item Have you read the ethics review guidelines and ensured that your paper conforms to them?
    \answerYes{Yes.}
\end{enumerate}

\item Additionally, if your study involves hypotheses testing...
\begin{enumerate}
  \item Did you clearly state the assumptions underlying all theoretical results?
    \answerNA{NA}
  \item Have you provided justifications for all theoretical results?
    \answerNA{NA}
  \item Did you discuss competing hypotheses or theories that might challenge or complement your theoretical results?
    \answerNA{NA}
  \item Have you considered alternative mechanisms or explanations that might account for the same outcomes observed in your study?
    \answerNA{NA}
  \item Did you address potential biases or limitations in your theoretical framework?
    \answerNA{NA}
  \item Have you related your theoretical results to the existing literature in social science?
    \answerYes{Yes.}
  \item Did you discuss the implications of your theoretical results for policy, practice, or further research in the social science domain?
    \answerYes{Yes.}
\end{enumerate}

\item Additionally, if you are including theoretical proofs...
\begin{enumerate}
  \item Did you state the full set of assumptions of all theoretical results?
    \answerNA{NA}
	\item Did you include complete proofs of all theoretical results?
    \answerNA{NA}
\end{enumerate}

\item Additionally, if you ran machine learning experiments...
\begin{enumerate}
  \item Did you include the code, data, and instructions needed to reproduce the main experimental results (either in the supplemental material or as a URL)?
    \answerYes{Yes}
  \item Did you specify all the training details (e.g., data splits, hyperparameters, how they were chosen)?
    \answerYes{Yes, see Appendix.} 
     \item Did you report error bars (e.g., with respect to the random seed after running experiments multiple times)?
    \answerNo{No, while we did run some experiments with different random seeds, however, for larger models due to computational resource limitation, we weren't able to do so.}
	\item Did you include the total amount of compute and the type of resources used (e.g., type of GPUs, internal cluster, or cloud provider)?
    \answerYes{Yes, see Appendix}
     \item Do you justify how the proposed evaluation is sufficient and appropriate to the claims made? 
    \answerYes{Yes.}
     \item Do you discuss what is ``the cost`` of misclassification and fault (in)tolerance?
    \answerYes{Yes, see Appendix.}
  
\end{enumerate}

\item Additionally, if you are using existing assets (e.g., code, data, models) or curating/releasing new assets...
\begin{enumerate}
  \item If your work uses existing assets, did you cite the creators?
    \answerYes{Yes, we have cited the models, datasets, and libraries. See also the Appendix}
  \item Did you mention the license of the assets?
    \answerNo{Yes, the models, datasets, and libraries are all licensed for research and non-profit use and are publicly available.}
  \item Did you include any new assets in the supplemental material or as a URL?
    \answerYes{Yes.} 
  \item Did you discuss whether and how consent was obtained from people whose data you're using/curating?
    \answerNA{NA, we use publicly available data.}
  \item Did you discuss whether the data you are using/curating contains personally identifiable information or offensive content?
    \answerYes{Yes, see the Discussion. }
\item If you are curating or releasing new datasets, did you discuss how you intend to make your datasets FAIR?
\answerYes{Yes, see the Discussion. }
\item If you are curating or releasing new datasets, did you create a Datasheet for the Dataset? 
\answerNo{We created the Hugging Face Hub~\footnote{\url{https://huggingface.co/docs/hub/en/index}} for models and datasets, making them publicly available. }
\end{enumerate}

\item Additionally, if you used crowdsourcing or conducted research with human subjects...
\begin{enumerate}
  \item Did you include the full text of instructions given to participants and screenshots?
    \answerNA{NA}
  \item Did you describe any potential participant risks, with mentions of Institutional Review Board (IRB) approvals?
    \answerNA{NA}
  \item Did you include the estimated hourly wage paid to participants and the total amount spent on participant compensation?
    \answerNA{NA}
   \item Did you discuss how data is stored, shared, and deidentified?
   \answerNA{NA}
\end{enumerate}

\end{enumerate}

\appendix
 \section*{Appendix}

\subsection{Data Preprocessing}
\label{sec:data_process_details}
The following preprocessing steps were applied to standardize and clean names:

\begin{itemize}
    \item \textbf{Removal of Extraneous Information:}  
    Nicknames, pronunciations, and metadata (e.g., content in parentheses, quotation marks, or following delimiters like commas or slashes) in Wikipedia introduction pages were removed. These elements are alternative spellings in other languages or pronunciation phonetics, and they introduce noise and are not part of the core name.

    \item \textbf{Normalization and Standardization:}  
    Punctuation (e.g., periods, commas, quotes) was replaced with spaces, and common prefixes (e.g., ``Dr'', ``Sir'') and suffixes (e.g., ``Jr'', ``III'') were removed.  
    Additionally, mojibake and Unicode problems were fixed by using the \texttt{ftfy\footnote{\url{https://ftfy.readthedocs.io/en/latest/}}} library, and Unicode normalization was performed using \texttt{unicodedata\footnote{\url{https://docs.python.org/3/library/unicodedata.html}}} to standardize encoding and ensure compatibility.

    \item \textbf{Validation and Cleaning:}  
    Names containing invalid characters (e.g., symbols, digits) were removed, and whitespace was standardized.  
\end{itemize}

During the preprocessing steps, only 0.15\% of the names were identified as invalid and removed, while all remaining non-English characters were converted to their corresponding English equivalents or removed.  
Table \ref{tab:preprocess_effect} illustrates some examples of the changes made during the process.

\begin{table}[!htp]\centering
\scriptsize
\begin{tabular}{lrr}\toprule
Before Preprocessing & After Preprocessing \\\midrule
JÃ¼rgen HÃ¼holdt & jurgen huholdt \\
RomÃ¡n MontaÃ±ez & roman montanez \\
Søren Hess-Olesen & sren hess olesen \\
Mónika Remsei & monika remsei \\
Rupert König & rupert konig \\
Miladin Ševarlić & miladin sevarlic \\
\bottomrule
\end{tabular}
\caption{Examples of names that were normalized and edited during preprocessing. }\label{tab:preprocess_effect}
\end{table}

For country names, we standardized all country names based on the ISO 3166 country codes\footnote{\url{https://www.iso.org/iso-3166-country-codes.html}}. These steps ensure the dataset contains clean and consistent names. 

\subsection{Rationale for Country Inclusion and Exclusion}
\label{countries_inclusion_exclusion}
To justify the inclusion and exclusion of multicultural countries, we used two metrics to guide our decisions: (1) Entropy, which measures the uniformity of character n-grams in a country's names, where higher values indicate greater naming diversity and overlap with other countries, and (2) F1 score, which reflects the model's classification performance, with lower scores indicating poorer performance.

Table \ref{tab:entropy} shows that excluded countries such as the USA, Canada, Australia, New Zealand, and South Africa have higher entropy values and lower F1 scores, which negatively impact the classification performance of other countries, particularly the UK. In contrast, included countries like the UK, Germany, and France exhibit lower entropy and higher F1 scores, justifying their retention. Based on these findings, we excluded South Africa and New Zealand due to their high entropy and low F1 scores, while retaining the UK to balance diversity with classification performance. Additionally, we needed one country to represent the English category in our classification and selected the UK for this task. (Normalized entropy is the scaled entropy value, adjusted to a range between 0 and 1.)

\begin{table*}[!htp]\centering
\scriptsize
\begin{tabular}{p{3.5cm}p{1cm}p{1cm}p{1cm}p{1cm}p{1cm}p{1cm}p{1cm}p{1cm}p{0.7cm}p{0.7cm}}\toprule
&United States &Canada &Australia &New Zealand &South Africa &United Kingdom &Germany &France &average \\\midrule
Normalized Entropy &0.96 &0.96 &0.94 &0.93 &1.00 &0.90 &0.86 &0.78 &0.72 \\
F1 score (Before excluding) &0.65 &0.34 &0.22 &0.23 &0.45 &0.54 &0.75 &0.73 &0.60 \\
F1 score (After excluding) &- &- &- &- &- &0.77 &0.78 &0.76 &0.63 \\
\bottomrule
\end{tabular}
\caption{Normalized entropy values and F1 scores (before and after exclusion) for a few countries.}\label{tab:entropy}
\end{table*}

\subsection{Training Details}
\label{sec:parameters_details}
We implemented our work in Python, using various libraries for different models (which are listed in a ``requirements.txt"). For FastText, we used its official Python library\footnote{\url{https://fasttext.cc/}} and trained it from scratch. For the neural networks (char-LSTM and char-CNN), we used PyTorch and TorchText. ByT5 was implemented using the Transformers and Datasets libraries. We used Optuna~\cite{akiba2019optuna} for hyperparameter tuning for all models except FastText, which has its own parameter-tuning function.

The best hyperparameters for char-CNN and char-LSTM include a learning rate of 0.005, weight decay of 1e-4, dropout of 0.2, embedding size of 100, and the AdamW optimizer~\citep{loshchilov2017decoupled} with a linear learning rate scheduler. For ByT5, the optimal learning rate is 3e-5, weight decay is 1e-7, and the AdamW optimizer is used without a learning rate scheduler. For FastText, we set the learning rate to 0.1, the embedding size to 100, and used an n-gram size of 5.
FastText and ByT5 have their own tokenizer and character encoder, while for char-LSTM and char-CNN, we generated character embeddings using TorchText with a maximum length of 64.

\subsection{Computational Resources}
\label{sec:computational_details}
We used ``NVIDIA GeForce GTX 1080 Ti'' GPU servers for our model training experiments. To accommodate the large models, we used ``gradient accumulation'' and ``mixed precision'' techniques. Gradient accumulation helps manage memory usage by accumulating gradients over several mini-batches before performing an update, while mixed precision training optimizes performance and reduces memory consumption by using both 16-bit floating-point numbers.


\subsection{Test Datasets}
\label{sec:test_datasets}

\begin{figure}[!ht]
    \centering
    \includegraphics[width=.45\textwidth]
    {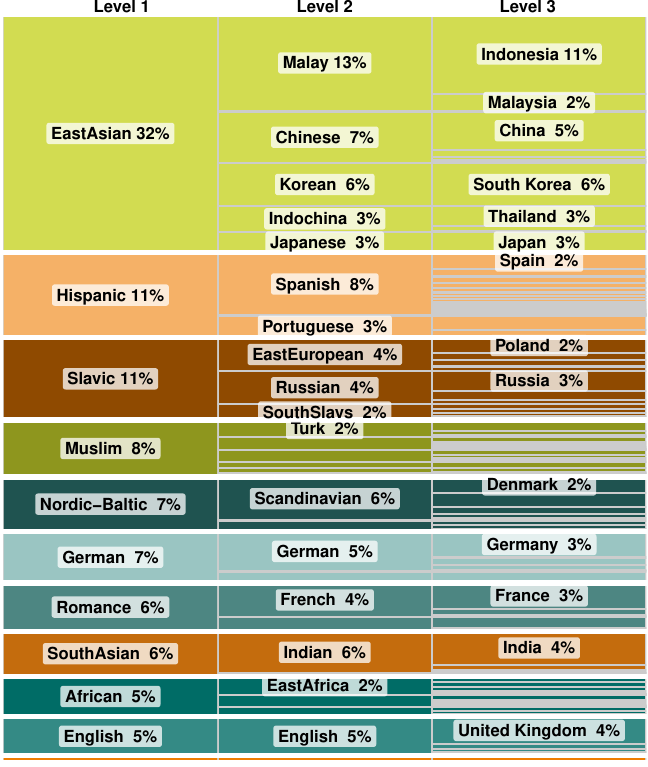}

    \includegraphics[width=.45\textwidth]
    {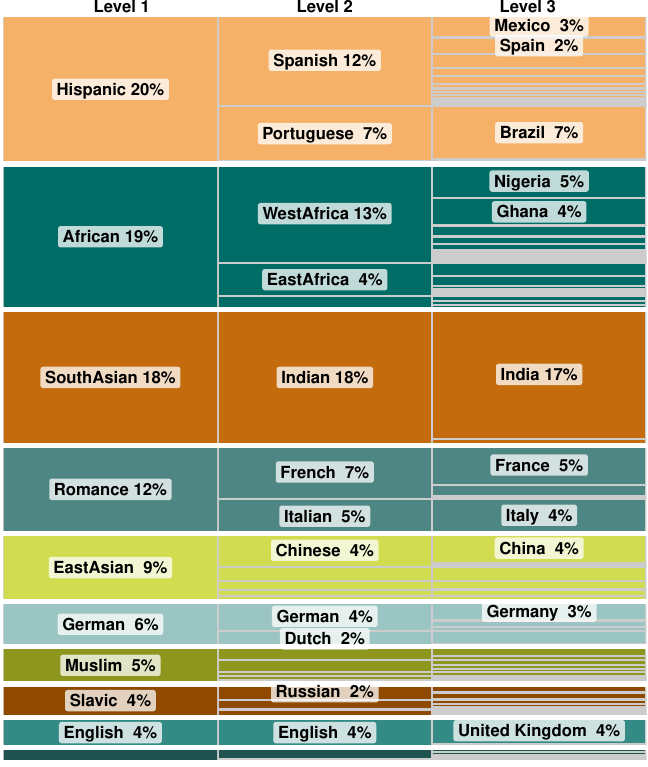}
    \caption{Distribution of nationality classes in the Athletes (top) and IUSSP (bottom) out-of-domain test datasets. Each column sums up to 100\% and represents one of the three classification levels, from the left (broader) to the right (more detailed). All classes are shown, but only those bigger than 2\% of the total are labeled.}
    \label{fig:treemap_test}
\end{figure}

Figure \ref{fig:treemap_test} shows the distribution of classes at each of the three levels for the Athletes (top) and IUSSP (bottom) datasets. At the first level, East Asian and Hispanic are the largest classes in the Athletes and IUSSP datasets, respectively. The class distribution in all three levels differs between the two test sets and is distinct from the Wikipedia data used to train and test the model. Despite the differences, the evaluation metrics are similar in all three datasets (presented in the main text), indicating the robustness of our model.  

\subsection{Error Analysis}
\label{error_analysis}
Here we clarify the errors and inaccuracies in our ML classification. We know that the ML models used have errors in classification and could still be improved, especially in the third level. However, we faced some misclassifications that could be due to other reasons. Below, we clarify examples of the observed errors. We explain potential reasons for these errors and also present our reasoning as to why the errors do not significantly affect our paper's conclusions:

\begin{itemize}    
    \item \textbf{Misclassification of names from countries with cultural/historical similarity}: In some cases, the classification models confuse names from culturally or linguistically similar countries. For example, the names listed in Table \ref{tab:example_belgium_netherlands} illustrate that Flemish individuals who, according to Wikipedia, have Belgian nationality might have names that are similar to those in the Netherlands. This could lead to the misclassification of their nationality to the Netherlands rather than Belgian, which is the true label based on Wikipedia data. Countries with historical or cultural relationships, such as Spain and Mexico, often have overlapping name structures, causing misclassifications. Another example is individuals of different nationalities who share the same name i.e., homonyms, such as ``Alexander Stein~\footnote{\url{https://de.wikipedia.org/wiki/Alexander\_Stein}}." There are four Wikipedia pages with this name for distinct individuals with various nationalities: German, American, Latvian, and Russian.
    \item \textbf{Second generation migrants with names from the country of origin or changed nationality}:
    The second generation migrants sometimes have names that are based on their parents' country of origin. They might obtain the citizenship of the host country, which is reflected in the Wikipedia introduction page, while they maintain the same names. This could lead to misclassifications in the nationality detection models, as shown in the examples in Table \ref{tab:second_generation}.
    \item \textbf{Error in datasets based on Wikipedia}:
    We combined multiple data sources collected from Wikipedia, which included the names and nationalities of individuals. Upon reviewing the data, we can assert that most of the name-nationality pairs were correct; however, minor errors did exist. On the one hand, the Nationality Estimate dataset \cite{le2021analysis} extracts information from Wikipedia descriptions to assign a country of origin to individuals whose pages lack explicit nationality details. However, this method can lead to errors when descriptions mention the country where the individual studied or lived instead of their actual country of origin. Table \ref{tab:error_in_data_1} illustrates examples of the inaccuracies in which the nationality was incorrectly stated in the data. On the other hand, about 50 percent of the nationality information in the Politicians dataset \cite{politicians} was missing. To address this, we used the DBpedia API to fill in the gaps. We checked if the pages contained nationality or country information, If it was not available, we used the birthplace of the politicians as a substitute for their nationality. While this approach helped solve problems, it also introduced some errors, as shown by the examples in Table \ref{tab:error_in_data_2}.    
    \item \textbf{Imbalances and classes with a small number of members}: 
    Due to the imbalanced nature of our training dataset (please see Figure \ref{fig:treemap} in the main text with the composition of classes in three levels), the model often predicts major classes when presented with data from classes with a small number of members, as it has not had sufficient exposure to learn to recognize them. 
\end{itemize}

    \begin{table}[!htp]\centering
    \scriptsize
    \begin{tabular}{lrrr}\toprule
    Name &Nationality in Wikipedia &Predicted nationality \\\midrule
    Bob van Laerhoven &Belgium &Netherlands \\
    Marc van den Abeelen &Belgium &Netherlands \\
    Jan van Hembyze &Belgium &Netherlands \\
    \bottomrule
    \end{tabular}
    \caption{Examples of misclassification of Flemish people with Belgian nationality to the Netherlands.}\label{tab:example_belgium_netherlands}
    \end{table}

    \begin{table}[!htp]\centering
    \scriptsize
    \begin{tabular}{lrrr}\toprule
    Name &Nationality in Wikipedia &Predicted nationality \\\midrule
    Bhavna Limbachia &United Kingdom &India \\
    Mahmoud Dahoud & Germany & Egypt\\
    Xu Pei &Germany &China \\
    \bottomrule
    \end{tabular}
    \caption{Misclassification of second-generation migrants due to names similar to country of origin or changes in nationality in Wikipedia.}\label{tab:second_generation}
    \end{table}

        \begin{table}[!htp]\centering
    \scriptsize
    \begin{tabular}{lrrr}\toprule
    Name & Nationality &Nationality in our dataset \\\midrule
    Arthur Glasser &USA &China \\
    Daniel Bayerstorfer &Germany &China \\
    Christopher Ludwig Eisgruber &USA &India \\
    \bottomrule
    \end{tabular}
    \caption{Errors in the training dataset because of incorrect nationality extraction from Wikipedia page descriptions.}\label{tab:error_in_data_1}
    \end{table}

    \begin{table}[!htp]\centering
    \scriptsize
    \begin{tabular}{lrrr}\toprule
    Name & Nationality &Birthplace\\\midrule
    Georg Heinrich Schnell &Germany &China \\
    Mathangi Arulpragasam &Sri Lanka &UK \\
    \bottomrule
    \end{tabular}
    \caption{Errors in the training dataset because of missing nationality information and using Birthplace instead.}\label{tab:error_in_data_2}
    \end{table}

Regarding the impact of the errors on our findings, we used a large corpus of 2.6 million name-nationality pairs, which ensured that the model was trained on diverse data, which is substantially larger than previous studies. Despite errors in certain country pairs, the model's performance remained consistent, as demonstrated by the stable results on both the training and out-of-domain test sets. 

Our main finding in the Scopus analysis predominantly highlights specific origin-destination country pairs (or migration corridors) with the highest number of scholars moving between them, such as the USA-China corridor. For these highly travelled corridors, our model has a very reliable performance. Additionally, our model demonstrated strong performance at the first and second levels. However, future studies aiming to differentiate names from culturally and linguistically similar countries, such as Spanish-speaking countries or Flemish regions, should incorporate additional information, as name-based nationality detection alone may not be sufficient.

\subsection{Assessment Per Class}
\label{sec:assessment_1}
\begin{figure}[!ht]
    \centering
    \includegraphics{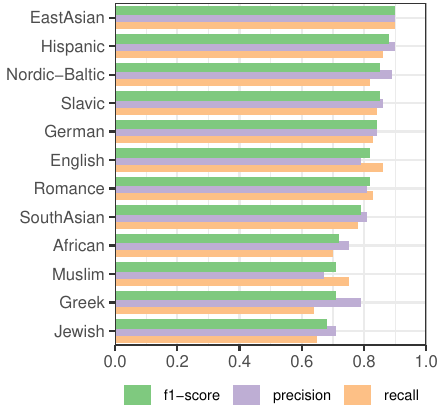}
    \caption{Evaluation metrics for the first level classification}
    \label{fig:evaluation_1}
\end{figure}

\begin{figure}[!ht]
    \centering
    \includegraphics
    {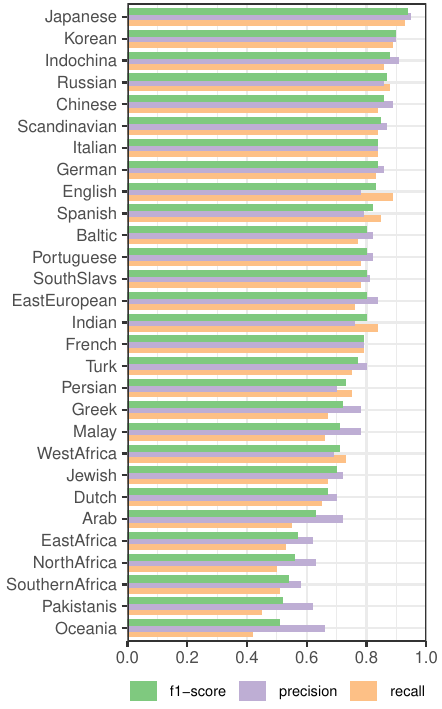}
    \caption{Evaluation metrics for the second level classification}
    \label{fig:evaluation_2}
\end{figure}

We examined the model's performance in classifying at levels 1 (see Figure \ref{fig:evaluation_1}) and 2 (see Figure \ref{fig:evaluation_2}). The assessment criteria used are precision, recall, and F1 score. It is evident that the model demonstrates higher precision compared to recall in both levels. This indicates that the model accurately predicts each class. Moreover, for classes with substantial data, such as European countries and regions, the average F1 score reached approximately 80\%. Conversely, classes with limited data have an F1 score of about 60\%. Although Korea, Indochina, and Greece have less data, they achieved a precision rate of around 80\%, demonstrating the model's ability to recognize names from these countries, perhaps due to distinct patterns in names from these countries. This assessment shows for which regions and groups of countries our model is the most reliable. 

\begin{figure}[!ht]
    \centering
    \includegraphics{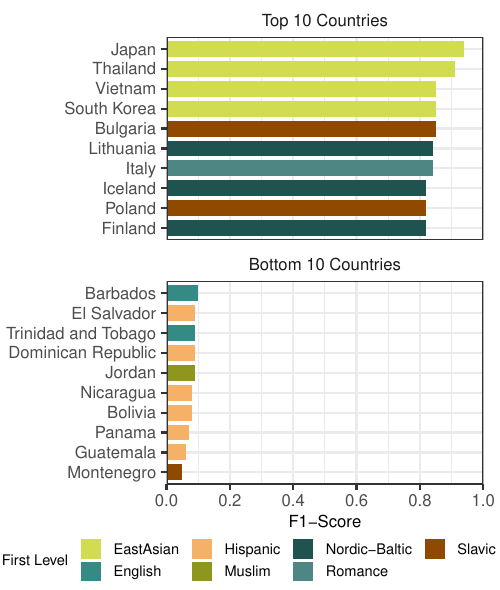}
    \caption{F1 score for the 10 countries with the highest scores (top) and 10 countries with the lowest scores (bottom). Only countries with more than 200 names in the Wikipedia test subset were considered.}
    \label{fig:evaluation_3}
\end{figure}

We also examined the countries for which our model is most and least reliable. Considering only countries with at least 200 names in the Wikipedia test subset, the F1 score ranges from 05\% for Montenegro to 94\% for Japan. This high variability across countries suggests that the model is appropriate for analyzing some migration corridors but may not be reliable for some origin and destination pairs. Figure \ref{fig:evaluation_3} shows the 10 countries with the highest and lowest F1 scores. The variability in the performance of the model at the country level exists even between countries within the same class at the first and second levels. While some first-level classes are only present in the top 10, such as East Asian with 4 out of 10 countries, or in the bottom 10, such as Hispanic with 6 out of 10, other countries with Slavic names appear in the top and bottom 10. 

\begin{figure*}[!ht]
    \centering
    \includegraphics[width=.95\textwidth]
    {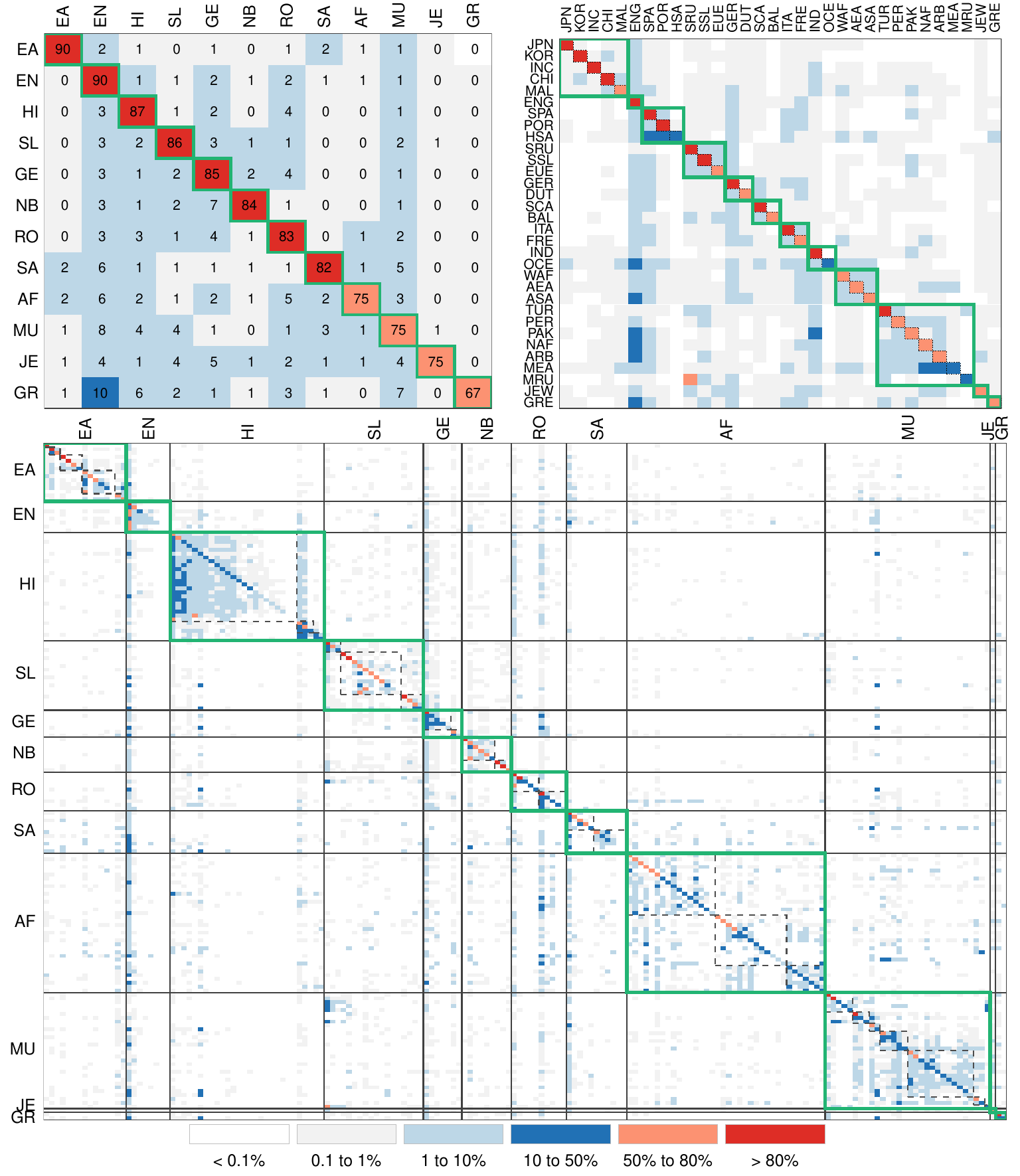}
    \caption{Confusion matrices for the first- (top left), second- (top right), and third-level classes (bottom). The rows represent the true and the columns show the predicted values. Colors represent the percentage of observations in each cell, and each row sums up to 100\%. In all matrices, the green rectangle highlights the cells for which the true value is equal to the predicted value at the first level, and the dashed rectangle highlights the classes at the second level. Labels used for the first level: African (AF), East Asian (EA), English (EN), German (GE), Greek (GR), Hispanic (HI), Jewish (JE), Muslim (MU), Nordic-Baltic (NB), Romance (RO), Slavic (SL), and South Asian (SA). Second level labels: Arab (ARB), Baltic (BAL), Chinese (CHI), Dutch (DUT), East Africa (EAF), East European (EUE), English (ENG), French (FRE), German (GER), Greek (GRE), Indian (IND), Indochina (INC), Japanese (JPN), Jewish (JEW), Korean (KOR), Malay (MAL), North Africa (NAF), Oceania (OCE), Pakistanis (PAK), Persian (PER), Portuguese (POR), Russian (RUS), Scandinavian (SCA), Southern Africa (SAF), South Slavs (SSL), Spanish (SPA), Turk (TUR), West Africa (WAF)}
\label{fig:confusion_matrices}
\end{figure*}

Figure \ref{fig:confusion_matrices} shows the confusion matrices for the first (top left), second (top right), and third levels (bottom). At the first level, about 87\% of the names for which the true value is Hispanic were predicted as Hispanic, showing that the model can identify names of Hispanic origin. At the second level for Hispanic names, the model correctly predicted more than 80\% of Spanish and Portuguese names, but the predicted values for countries with Hispanic Southern African names, i.e., Angola and Mozambique, are distributed across Spanish, Portuguese, and Hispanic Southern African classes. 

Our model can correctly identify a name with a Hispanic pattern and even distinguish between Spanish and Portuguese names, but the model cannot correctly identify the country within the second-level class. The same applies to English names; the model can determine the English origin but not the specific country. In both examples, the model cannot correctly identify the countries because they have similar names, and there are imbalances and classes with a small number of cases at the country level. On the one hand, the misclassification of names from countries with cultural/historical similarity leads to a spread of predicted values across multiple countries; this pattern is clear for most of the Spanish countries in the country-level confusion matrix (see the first rows of the bottom matrix in Figure \ref{fig:confusion_matrices}). On the other hand, the imbalance and minor classes lead the model to predict the same country for different true values, and other countries are not predicted even when the name is from that country. (See the first and last two columns of the Hispanic class in the bottom matrix in Figure \ref{fig:confusion_matrices}). 

\clearpage

\end{document}